\documentclass{aa} 
\usepackage{txfonts}
\usepackage{adjustbox}
\usepackage{makecell}
\usepackage[colorlinks,citecolor=blue, linkcolor=blue]{hyperref}
\usepackage{graphicx}
\usepackage{amsmath}
\usepackage{ifoddpage}
\usepackage{listings}
\usepackage{xcolor}
\usepackage{natbib}
\usepackage{appendix}
\usepackage{placeins}
\bibpunct{(}{)}{;}{a}{}{,}

\begin{document}

\title{Large-scale variability in macroturbulence driven by pulsations in the rapidly rotating massive star $\zeta$~Oph from high-cadence ESPRESSO spectroscopy and TESS photometry\thanks{Based on observations collected at the European Southern Observatory under ESO programme 113.26B9.}}

\titlerunning{Large-scale variability in macroturbulence in $\zeta$~Oph}

\author{A. J. Kalita\inst{1},
          D. M. Bowman\inst{1,2},
          M. Abdul-Masih\inst{3,4},
          S. Sim{\'o}n-D{\'i}az\inst{3,4}
          }

\authorrunning{Kalita et al.}

   \institute{$^1$ School of Mathematics, Statistics and Physics, Newcastle University, Newcastle upon Tyne, NE1 7RU, United Kingdom
              \\
              \email{a.j.kalita2@newcastle.ac.uk
              }
              \\
              $^2$ Institute of Astronomy, KU Leuven, Celestijnenlaan 200D, 3001 Leuven, Belgium
              \\
              $^3$ Instituto de Astrof{\'i}sica de Canarias, C. Vıa Lactea, s/n, 38205 La Laguna, Santa Cruz de Tenerife, Spain
              \\
              $^4$ Universidad de La Laguna, Dpto. Astrof{\'i}sica, Av. Astrof{\'i}sico Francisco Sanchez, 38206 La Laguna, Santa Cruz de Tenerife, Spain}
    \date{Received YYYY; accepted: YYYY}
 
    \abstract{Despite their importance, the dynamical properties of massive stars remain poorly understood. Rotation is a key driver of internal mixing and angular momentum transport, significantly influencing massive star evolution. However, constraining rotation from spectroscopy is challenging, as spectral lines often exhibit excess broadening beyond rotation. The origin of this additional broadening, typically attributed to large-scale velocity fields and commonly referred to as macroturbulence, remains uncertain and unconstrained. Here, we present the combined analysis of TESS photometry and rapid time-series spectroscopy using the high-resolution ESPRESSO instrument at the Very Large Telescope of the European Southern Observatory for the rapidly rotating and pulsating massive star $\zeta$~Oph. Leveraging excellent temporal coverage, our analysis demonstrates that pulsation-induced variability leads to peak-to-peak scatter as large as 88~km\,s$^{-1}$ in the observed macroturbulent velocity time series. We also demonstrate that time-averaged macroturbulent velocities are spectral line specific and can exceed 100~km\,s$^{-1}$. Furthermore, the macroturbulent velocities from shorter integration times are systematically lower than those derived from stacked spectra mimicking longer exposure times typically needed for fainter stars. These results highlight the role of pulsations in driving variable macroturbulence in massive stars, while also pointing to a potential bias in spectroscopic estimates of macroturbulence for fainter massive stars.}


    \maketitle

\section{Introduction}
\label{sec:introduction}

Massive stars are progenitors of neutron stars and black holes, and play a central role in galactic chemical evolution and star formation \citep{kippenhan_1990, Bromm_first_stars2004}. Constraining their evolution is key to understanding both compact object populations and galaxy evolution. However, several critical processes remain poorly understood. Rotation, in particular, influences stellar evolution through internal mixing and angular momentum transport \citep{meynet_and_maeder_review_2000, norbert_langer_2012, conny_tamara_rogers_2019}, but its measurement via spectroscopy is complicated by an additional source of broadening known as macroturbulence.

Macroturbulence is often attributed to large-scale velocity fields in stellar atmospheres, though its physical origin remains poorly understood and unconstrained \citep{aerts2009collective, grassitelli2015apj, simon_iacob_2017}. Three main hypotheses have been proposed to explain the large macroturbulent velocities observed in massive stars. \citet{aerts2009collective} suggested that the collective effect of numerous low-amplitude gravity-mode pulsations can give rise to the additional broadening of the spectral line profile, which is usually attributed to macroturbulence. \citet{grassitelli2015apj} identified a correlation between macroturbulence and turbulent pressure derived from 1D stellar structure models, proposing that pressure fluctuations in sub-surface convection zones can drive small-amplitude pulsations, which in turn result in the observed velocity fields. More recently, \citet{dwaipayan2024}, using 2D hydrodynamic simulations, demonstrated that gas pockets near the iron-opacity bump can be radiatively accelerated to velocities comparable to those needed for macroturbulence. Previous spectroscopic studies have suggested that the time-averaged effects of pulsations generates the velocity field responsible for macroturbulence (i.e.\ $v_{\rm macro}$) in massive stars \citep{aerts2009collective, aerts2014use, conny_tami_convective_IGW_2015}. Moreover, massive stars typically have large values of $v_{\rm macro}$ and largely located within theoretical pulsation instability regions in the Hertzsprung--Russell (HR) diagram \citep{simon_iacob_2017, Godart2017, Burssens2020}. However, the time-dependent relationship between macroturbulence and pulsations, as well as the impact on inferred atmospheric parameters, remains unclear.

To date, there are only a handful of studies that have investigated line profile variability (LPV) in massive stars and its connection to multiperiodic pulsations and macroturbulence (see \citealt{bowman2020asteroseismology} for a review). One example is the nearest O-type star, $\zeta$~Oph \citep{kambe1997multiperiodicity, walker2005pulsations, howarth2014amplitude}. Time-series high-resolution spectroscopy from various campaigns has allowed a dominant period of $3.33$~h (i.e.\ $\nu = 7.18$~d$^{-1}$) to be consistently observed and attributed to a non-radial pulsation mode with spherical harmonic indices $\ell = |m| = 4$ \citep{kambe1990spectroscopic, reid1993time, kambe1997multiperiodicity, balona1999moving}. Other reported pulsation periods include $2.01$~h ($\nu = 11.89$~d$^{-1}$, $\ell = 8$; \citealt{balona1999moving, kambe1997multiperiodicity}), $5.34$~h \citep{kambe1997multiperiodicity}, $4.468$~h ($\nu = 5.37$~d$^{-1}$), and $2.43$~h ($\nu = 9.86$~d$^{-1}$; \citealt{reid1993time}). However, none of these investigations explored how such pulsations are related to the time-dependent nature of macroturbulence.

Space-based photometry from the Microvariability and Oscillations of Stars (MOST; \citealt{Walker2003MOST}) satellite further confirmed the multiperiodic pulsations of $\zeta$~Oph, detecting a persistent signal with a period of $4.63~\mathrm{h}$ (i.e. $\nu = 5.18~\mathrm{d^{-1}}$) across several years \citep{walker2005pulsations}. Using SMEI, MOST, and WIRE space photometry, \citet{howarth2014amplitude} corroborated the detection of the $4.63$~h and $3.33$~h periods and revealed additional pulsations with periods of $8.11$~h ($\nu = 2.96$~d$^{-1}$) and $8.99$~h ($\nu = 2.67$~d$^{-1}$). Therefore, it is well established that $\zeta$~Oph is a multiperiodic pulsator, but that its variability is also not strictly periodic \citep{howarth2014amplitude}. Moreover, the connection between its photometric and spectroscopic pulsation periods warrants further investigation.

In this paper, we analyse the pulsations and various diagnostics of LPV, including macroturbulence, for $\zeta$~Oph using new TESS photometry and high-resolution spectroscopy from ESO's ESPRESSO spectrograph. $\zeta$~Oph is a rapidly rotating, pulsating O9.2\,IVnn star \citep{Sota_southern_stars_2014} with an uncertain projected rotational velocity of $320< v\,\sin\,i< 400$~km\,s$^{-1}$ in the literature \citep{conti1977spectroscopic, herrero1992intrinsic, jankov2000fourier, simon2014iacob}, which corresponds to a rotation period between 0.7 and 1.2~d. This uncertainty arises primarily because sparsely sampled spectra cannot disentangle the time-dependent contributions of rotation, pulsations, and macroturbulence in the line profiles. With unprecedented temporal coverage, we empirically measure the amplitudes and time scales in the line-specific variability in radial velocity, rotation rates, and macroturbulence induced by pulsations. We also investigate the impact of variable macroturbulence on the inferred stellar parameters, and potential biases in reported values of macroturbulence from the longer integration times needed for massive stars fainter than $\zeta$~Oph.

\section{Data description and reduction}
\label{sec: Data description}

Our spectroscopic data were obtained with the ESPRESSO spectrograph \citep{megevand2014espresso, pepe2021espresso} at the combined Coud{\'e} facility of the Very Large Telescope (VLT) of ESO, Chile, under programme 113.26B9 (PI: Bowman). All observations were conducted in the Ultra High-Resolution 1-UT (UHR) mode, providing a median resolving power of 190\,000 and a wavelength coverage of 380-788 nm. This configuration ensures a radial velocity (RV) precision of $\sim$5~m\,s$^{-1}$. A total of 1706 spectra were assembled, each with a 30-s exposure time over four consecutive nights of $7-10$ June 2024. The ESPRESSO spectra were reduced using the ESOReflex pipeline\footnote{ESO Reflex Pipeline: \href{https://www.eso.org/sci/software/esoreflex/}{www.eso.org/sci/software/esoreflex}} \citep{freudling2013esoreflex}. The reduced spectra were manually normalized using spline fitting to the continuum level. The mean signal-to-noise ratio (S/N) across all spectra at the wavelength of 550 nm is 393, with the 16th and 84th quantiles for S/N being 295 and 495, respectively.

The ESPRESSO spectra of $\zeta$~Oph are rich in metal lines, but given its rapid rotation rate \citep{herrero1992intrinsic, howarth2001rotational, simon2014iacob} most metal lines exhibit significant blending. However, the He\,{\sc i+ \sc ii} $\lambda$4026, He\,{\sc i} $\lambda$4143, He\,{\sc ii} $\lambda$4200, He\,{\sc i} $\lambda$4471, and He\,{\sc i} $\lambda$4922 are sufficiently isolated to perform a LPV study unaffected by blending. The Balmer lines are strongly affected by Stark broadening and are less sensitive to non-radial pulsations than helium lines. For these reasons, it is typical to focus on isolated helium lines in LPV studies of rapidly rotating massive stars (e.g. \citealt{balona1999moving}). In this paper, we show results for the He\,{\sc i+ \sc ii} $\lambda$4026, He\,{\sc ii} $\lambda$4200, and He\,{\sc i} $\lambda$4922 lines, and investigate how rotation, pulsations and macroturbulence vary among different ionisation stages of helium. We leave the analysis of blended metal lines for future work.

\section{Line profile variability analysis}
\label{section:lpv_analysis}

\begin{figure}
    \centering
    \includegraphics[width=0.99\columnwidth]{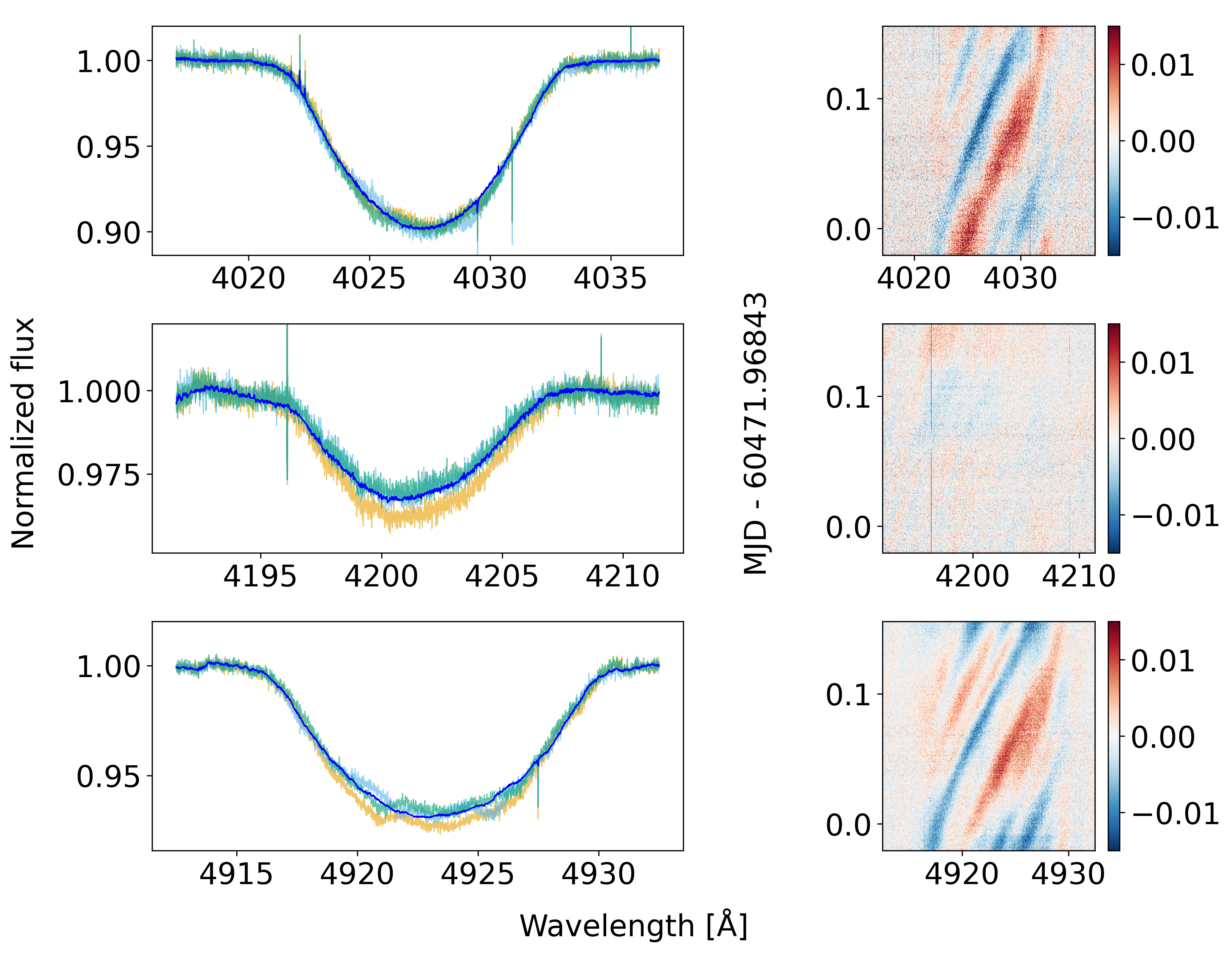}
    \caption{Left panels: Three spectra from 9 June 2024 showing LPV for the He\,{\sc i+\sc ii} $\lambda$4026, He\,{\sc ii} $\lambda$4200, and He\,{\sc i} $\lambda$4922 lines from top to bottom, respectively, with the stacked mean spectrum overplotted in dark blue. Right panels: Residual dynamic spectra for the corresponding left panels.}
    \label{lpv}
\end{figure}

Our ESPRESSO time-series of $\zeta$~Oph reveals significant LPV caused by multiperiodic pulsations. In Fig.~\ref{lpv}, we show three spectra selected from 9 June 2024 to demonstrate the variability in three helium lines, as well as the mean spectrum obtained by averaging all 1706 spectra overplotted in dark blue for each line. Figure~\ref{lpv} also shows residual dynamic spectra, constructed by subtracting the mean spectrum from each individual spectrum. The residual spectra are time-stacked along the y-axis and reveal LPV (i.e. `bumps' moving across the line profile) caused by non-radial pulsations (see e.g. \citealt{aerts2009collective}). In the following, we demonstrate how pulsation-induced LPV translates into variability in other spectroscopic diagnostics.

\subsection{Radial Velocity}\label{section:rv}

\begin{figure}
    \centering
    \includegraphics[width=0.99\columnwidth]{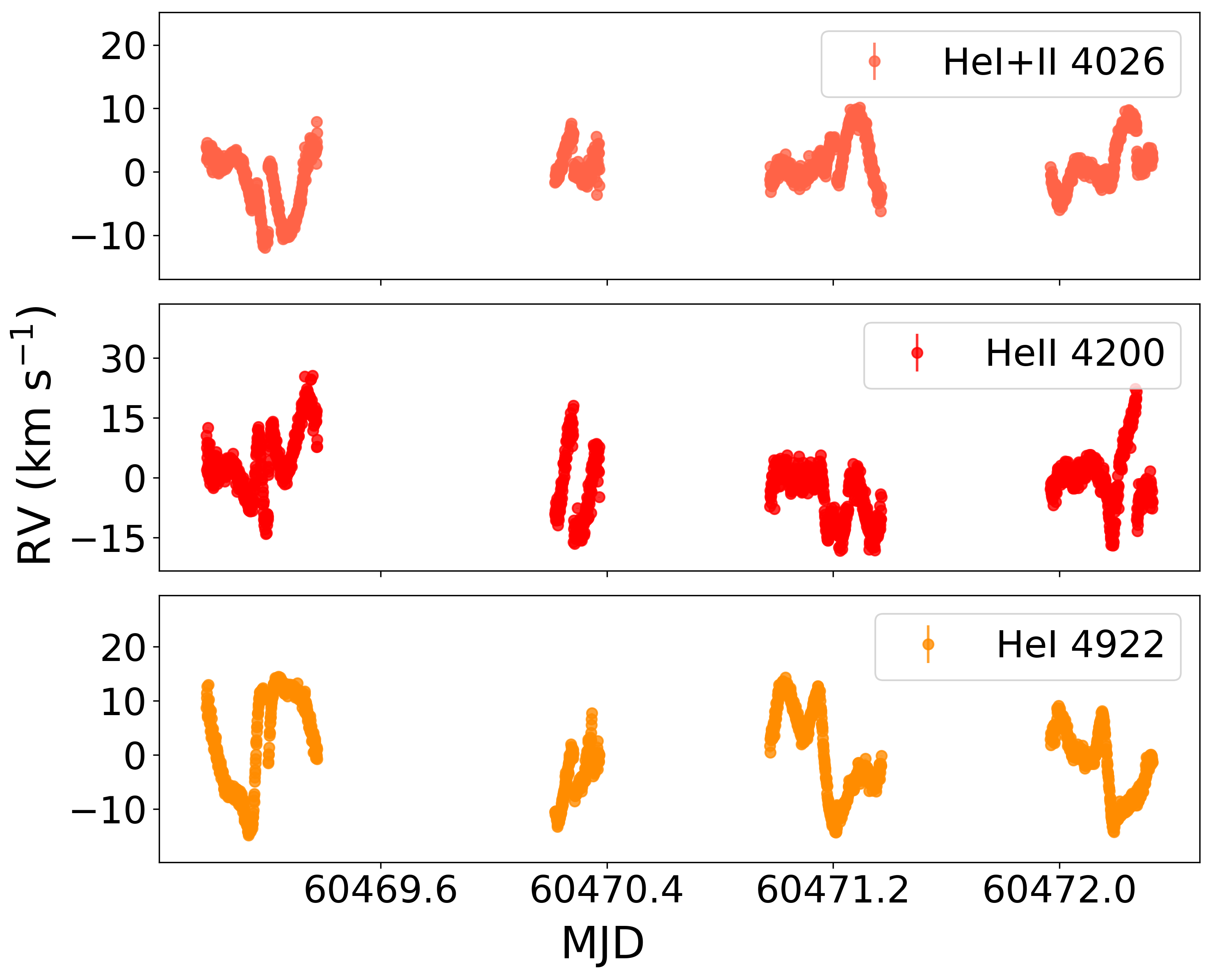}
    \caption{RV time series for three helium lines in $\zeta$~Oph extracted using cross correlation. The peak-to-peak RV variations exceed 20~km\,s$^{-1}$ in a multi-periodic manner because of pulsations rather than binarity.}
    \label{rv_timeseries}
\end{figure}

The RV stability of individual spectral lines was examined using the cross-correlation technique \citep{Zucker2003ccf, shenarLB1}. Figure~\ref{rv_timeseries} shows the observed RV variability for the same spectral lines shown in Fig.~\ref{lpv}, which all exhibit peak-to-peak RV variations exceeding 20~km\,s$^{-1}$. We investigated the dominant variability periods in the RV time series using an iterative pre-whitening technique \citep[see][]{bowman2021towards}. This method involves fitting a sinusoidal model of the form

\begin{equation}  
f(t) = \sum_{i = 1}^{n} A_i \sin(\omega_i (t - t_0) + \phi_i),
\label{eq:prewitenning}
\end{equation}
\noindent where $A_i$ is the amplitude, $\omega_i$ is the frequency identified from the Lomb-Scargle periodogram analysis of the time series (cf. \citealt{Scargle1982}), $t$ is the time, and $\phi_i$ is the phase relative to the reference time $t_0$, chosen as MJD = 60468.983772, which is the time stamp of the first spectrum. The pre-whitening process was continued until the S/N of the next highest peak in the frequency spectrum dropped below a false alarm probability of $0.001\%$. At each step, we carefully examined the corresponding spectral window of the ESPRESSO time series (see Fig.~\ref{spec_window}) to mitigate aliasing. After all significant periods had been identified and extracted, they were included in the model in Eqn.~(\ref{eq:prewitenning}) and optimised using least-squares fitting. This method is typical for analysing light curves of pulsating massive stars (see review by \citealt{bowman2020asteroseismology}).

Iterative pre-whitening of the RV times series of $\zeta$~Oph revealed a range of variability time scales and amplitudes, some of which were reported in previous studies. We provide the significant periods and their corresponding uncertainties from the RV (and other) time series analysis in this work in Table~\ref{tab:frequencies}. The 3.33-h period was observed for all three helium lines, which are shown in Fig.~\ref{rv_timeseries}. In the He\,{\sc ii} $\lambda$4200 and He\,{\sc i} $\lambda$4922 lines, the 3.33-h period was the dominant source of variability with an amplitude of 3~km\,s$^{-1}$, while for the He\,{\sc i+ \sc ii} $\lambda$4026 line, this was the second largest source of variability with an amplitude of 2~km\,s$^{-1}$. This period was previously shown to be due to a non-radial pulsation mode with spherical harmonic geometry $\ell = |m| = 4$ \citep{kambe1990spectroscopic, reid1993time, kambe1997multiperiodicity}. We also found a significant period of 2.04\,h in both the He\,{\sc i + ii} $\lambda$4026 and He\,{\sc i} $\lambda$4922 lines, while the He\,{\sc ii} $\lambda$4200 line exhibits a similar period of 2.00\,h. Additionally, we detect periods of 2.46\,h and 2.50\,h in the He\,{\sc ii} $\lambda$4026 and He\,{\sc i} $\lambda$4200 lines, respectively, which are close to the reported 2.43-h pulsation period measured by \citet{reid1993time} and \citet{kambe_short_term_lpv1993}. Later, \citet{kambe1997multiperiodicity} argued that the 2.43-h period could be an alias of the true pulsation period of 2.01~h. In our ESPRESSO RV time series data both of these periods are significant. However, in the TESS light curve of $\zeta$~Oph, we only find the 2.43-h period to be significant, while the 2.01-h period is below the significance threshold of 0.001\% false alarm probability (see Section~\ref{sec:TESS_photometry}). Thus, we conclude that the 2.43-h period is likely the real pulsation mode and the 2.01-h period is an alias, as concluded by \citet{kambe1997multiperiodicity}.

It is common for massive stars to be found in binaries, and the majority of them to interact with a companion during their lifetime \citep{sana_binary_interaction_2012, selma_de_mink_rotation_rates_2013, pablo_marchent_julia_annual_review_2024}. Therefore, we also searched for evidence of binarity in our RV time series. The high surface abundance of nitrogen and helium, along with the rapid surface rotation suggest that $\zeta$~Oph is a binary interaction product \citep{herrero1992intrinsic, villamariz_chemical_composition_2005}. Additionally, $\zeta$~Oph is also classified as a runaway star because of its high peculiar spatial velocity of 30~km\,s$^{-1}$ \citep{blaauw_runaway_paper_1961}, suggesting that it could have been ejected following the supernova explosion of its former primary companion. Based on this premise, \citet{mathieu_renzo_ylva_gotberg_zeta_oph_2021} showed that an accretor star with an initial mass of 17~M$_\odot$ in a 100-d binary orbit with a 25-M$_\odot$ donor can broadly reproduce the observed mass, radius, runaway velocity, surface rotation and surface abundances of helium, nitrogen and carbon of $\zeta$~Oph. Therefore, investigating any potential signatures of binarity for $\zeta$~Oph remains an important complementary goal of our work.

In most spectroscopic studies of massive stars that focus on binarity detection, pulsations are often treated as a noise term in the RV time series. A peak-to-peak RV amplitude exceeding 20~km\,s$^{-1}$ is typically interpreted as evidence for binarity \citep{sana2013vlt, dunstall2015vlt, laurent2022multiplicity, bloem2025julia}. However, we find significant RV variability above this threshold that is clearly driven by pulsations. Moreover, the variability is line-dependent and multi-periodic, consistent with earlier findings \citep{kambe1990spectroscopic, kambe1997multiperiodicity, walker2005pulsations, howarth2014amplitude}. Our results confirm the quasi-periodic nature of the pulsations: the recovered periods are close to, but not exactly the same as those previously reported. We therefore conclude that the observed RV variability in $\zeta$~Oph solely arises from pulsations, as we find no spectroscopic evidence for binarity in the RV time series.

Of course, given the observing cadence and time span of our ESPRESSO time series, we are most sensitive to the presence of a close binary companion. Using the mass of $\zeta$~Oph reported in literature (20.2 M$_\odot$; \citealt{repolust_wind_params_2004}) and its observed RV semi-amplitude from our ESPRESSO data (i.e. $\sim$15~km\,s$^{-1}$), we should be able to detect signatures of any binary companion of mass roughly 1~M$_\odot$ or larger from our RV time series. However, since the total observing window is only 4~d, we can only exclude the possibility of companions with periods less than about 10~d. Finally, despite the lack of binarity in $\zeta$~Oph, our results demonstrate the importance of accounting for LPV caused by pulsations in multiplicity studies of massive stars \citep[see][]{simon2024}.

\subsection{Rotation}\label{sec:rotation}

\begin{figure}
    \centering
    \includegraphics[width=0.99\columnwidth]{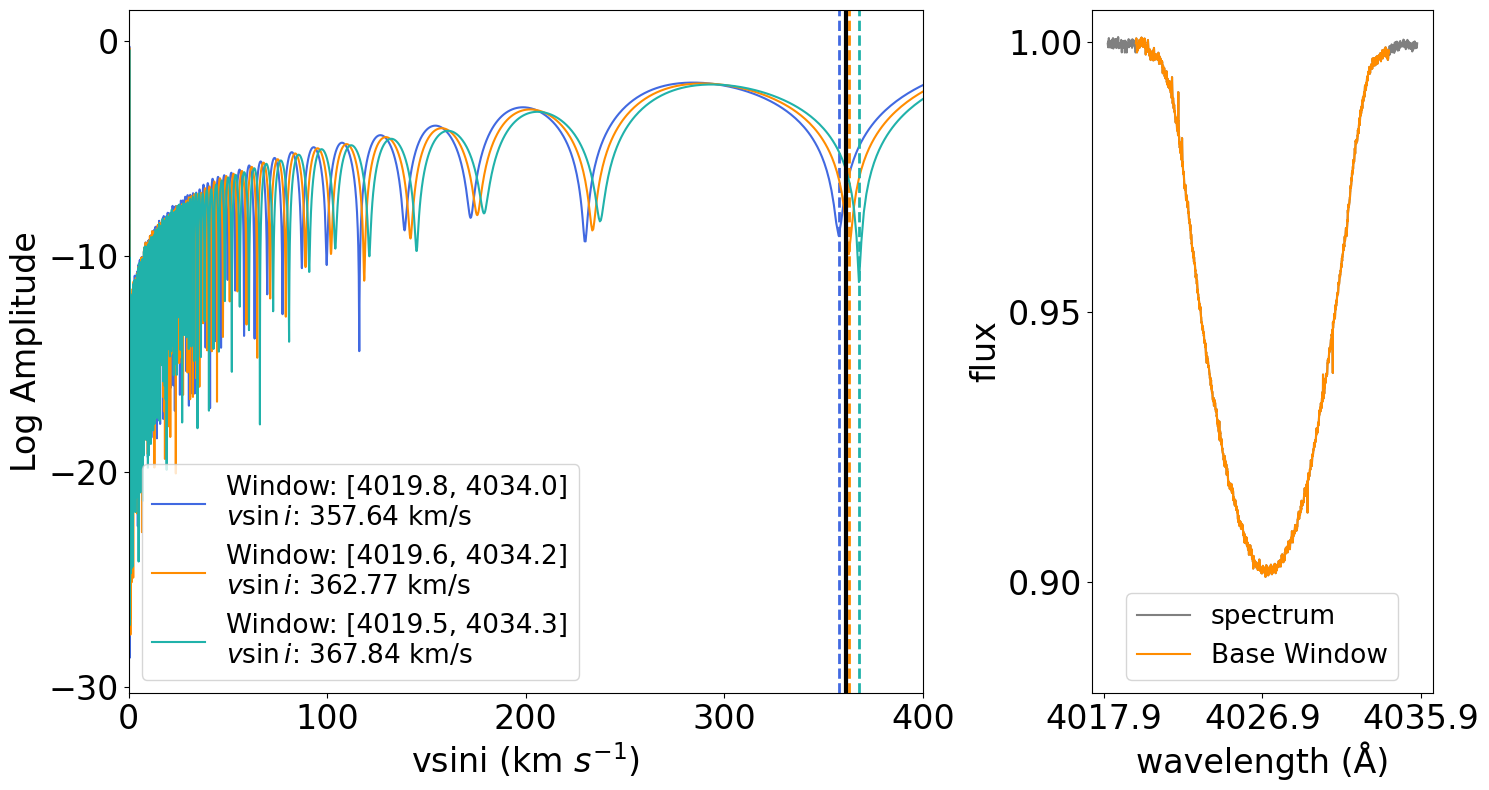}
    \caption{Left panel: Fourier transform of the line profile for different wavelength windows. The black line corresponds to the first minimum of the Fourier transform for the wavelength window shown in the right-hand panel. The coloured vertical dashed lines represent the first minimum of the Fourier transform obtained by adjusting the base window by $\pm0.1$\AA~ the line profile. Right panel: line profile for the He\,{\sc i+ \sc ii} $\lambda$4026 line in the stacked spectrum, with the selected wavelength window highlighted in orange.}
    \label{ft_lineprofile}
\end{figure}

Determining the precise projected rotational velocity ($v\,\sin\,i$) of a rapidly rotating massive star such as $\zeta$~Oph can be challenging due to the complexities caused by LPV. \citet{conti1977spectroscopic} estimated a $v\,\sin\,i$ of 351~km\,s$^{-1}$ using spectra obtained at the Cerro Tololo Observatory and comparing with the non-LTE atmosphere models. \citet{penny1996projected} reported a value of 348~km\,s$^{-1}$ based on cross-correlation with narrow-lined O-star line-list templates. \citet{howarth2001rotational} derived a $v\,\sin\,i$ of 400~km\,s$^{-1}$ using non-LTE atmospheric model fitting, which is consistent with earlier estimates by \citet{herrero1992intrinsic} and \citet{balona1999moving}. \citet{jankov2000fourier} applied the Fourier technique to an averaged spectrum from 242 epochs assembled at the Laboratorio Nacional de Astrofisica, reporting $v\,\sin\,i$ = 390~km\,s$^{-1}$. However, none of these studies accounted for (time-variable) macroturbulence in the broadening estimation of the spectra. The most recent estimate of 319~km\,s$^{-1}$ by \cite{simon2014iacob} and 385~km\,s$^{-1}$ by \cite{holgado_rotation_rates_for_Ostars_2022} combines the Fourier method with line-profile fitting, explicitly accounting for macroturbulence, but did not consider its time-dependent nature owing to the limited number of epochs.

We determine the $v\,\sin\,i$ of $\zeta$~Oph using the Fourier transform method, following the approach in which the first zero of the Fourier transform of a spectral line profile corresponds to $v\,\sin\,i$ for stars dominated by rotational broadening \citep{gray2005observation, simon2007fourier, simon2014iacob}. While generally reliable and a widely used approach, this method can introduce scatter in the measurements, particularly in stars with significant pulsational broadening and/or LPV, which is an important consideration in the spectroscopic analysis of massive stars (see e.g. \citealt{aerts2009collective, aerts2014use, simon_iacob_2017}). For stars with significant LPV, this method can lead to uncertainties of up to 5–10\% in the inferred $v\,\sin\,i$ \citep{piters_fourier_bessel_1996, pjgroot_fourier_bessel_1996}. The uncertainty increases if additional broadening, comparable in magnitude to the rotational broadening, is present in the line profile. On the other hand, for fast rotators such as $\zeta$~Oph, where rotational broadening dominates and often masks pulsation-induced asymmetries in the line profile, the Fourier method is considered reliable \citep{simon2007fourier, simon_iacob_2017}.

\begin{figure}
    \centering
    \includegraphics[width=0.99\columnwidth]{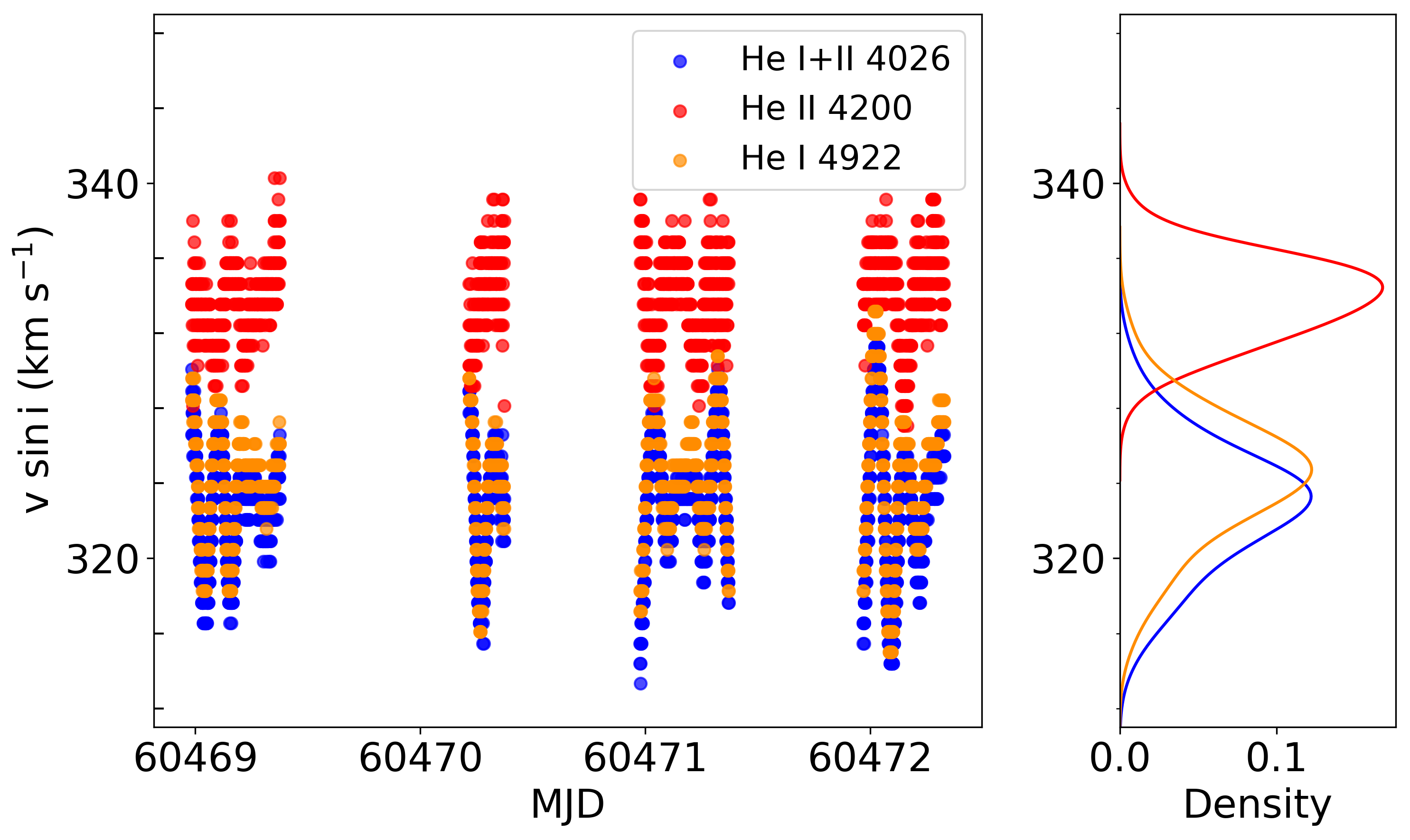}
    \caption{Left panel: $v\,\sin\,i$ time series for different spectral lines with peak-to-peak scatter up to 15~km\,s$^{-1}$. Right panel: Kernel density estimation of the scatter used to estimate the formal uncertainty in $v\,\sin\,i$.}
    \label{vsini_time-series}
\end{figure}

As an example of the application of the Fourier method to our ESPRESSO data, Fig.~\ref{ft_lineprofile} illustrates the Fourier transform of the He\,{\sc i} $\lambda$4922 line calculated based on the average spectrum of stacking all individual epochs. The line profile is initially selected within a base wavelength window as shown in Fig.~\ref{ft_lineprofile}, then systematically varied by $\pm 0.1$\AA~ at each wing to probe how much our results are sensitive to selection biases and normalisation. The corresponding $v\,\sin\,i$ values for each iteration are indicated by vertical dashed lines in Fig.~\ref{ft_lineprofile}. To ensure consistency in our subsequent analysis, the line region is manually identified in the stacked spectrum and fixed across all epochs before applying the Fourier transform to derive $v\,\sin\,i$. The bootstrapped uncertainty on $v\,\sin\,i$ measurement from individual epochs is systematically less than 2~km\,s$^{-1}$. However, as discussed previously, an uncertainty of $\leq 5$~km\,s$^{-1}$ (i.e. $\sim3\%$) is likely underestimated for a star with significant LPV such as $\zeta$~Oph (see \citealt{piters_fourier_bessel_1996}). The peak-to-peak scatter in the $v\,\sin\,i$ time series is shown in Fig.~\ref{vsini_time-series} for the same three helium lines and is of order 15~km\,s$^{-1}$ (i.e. 5\% of the mean $v\,\sin\,i$ value). This scatter can be treated as the lower limit for the uncertainty on $v\,\sin\,i$ from the Fourier method for a massive star with LPV such as $\zeta$~Oph. This is in agreement with the findings of \citet{piters_fourier_bessel_1996}, who suggested that the Fourier method for stars with LPV can yield up to 10\% uncertainty of the inferred $v\,\sin\,i$. Hence, $15-30$~km\,s$^{-1}$ (i.e. 5-10\%) is likely a more realistic uncertainty for our $v\,\sin\,i$ measurements.

Figure~\ref{vsini_time-series} shows the $v\,\sin\,i$ time series for three different spectral lines, He\,{\sc i+ \sc ii} $\lambda$4026, He\,{\sc ii} $\lambda$4200, and He\,{\sc i} $\lambda$4922, yielding mean values of $324 \pm 13$ km\,s$^{-1}$, $334 \pm 9$ km\,s$^{-1}$, and $322 \pm 13$ km\,s$^{-1}$, respectively. The indicative formal errors on $v\,\sin\,i$ from our Fourier analysis for each line are calculated as the 3-$\sigma$ confidence intervals of the kernel density based on all individual measurements (see Fig.~\ref{vsini_time-series}), thus accounting for the epoch-to-epoch scatter due to LPV. Again, we stress that these errors might be underestimated, and an uncertainty at a 10\% level might be a more reasonable error estimate for stars with LPV following the arguments described above. Interestingly, we find that the $v\,\sin\,i$ measured from the He\,{\sc ii} $\lambda$4200 line is systematically larger by about 10~km\,s$^{-1}$ compared to the other two lines. The difference in mean $v\,\sin\,i$ values across the three helium lines could be related to the limitations of the Fourier method as discussed previously. We note that further investigation is necessary to determine the cause of this discrepancy.

We applied our iterative pre-whitening method to the $v\,\sin\,i$ time series of $\zeta$~Oph and detected the 3.33-h, 2.43-h and 4.63-h periods amongst all three helium lines. As previously mentioned, the 2.43-h and 3.33-h periods are known non-radial pulsation modes, while the 4.63-h period is suggested to be either a radial first overtone or a low-angular degree non-radial mode \citep{kambe1990spectroscopic, kambe1997multiperiodicity, balona1999moving, walker2005pulsations}. Additionally, all three helium lines exhibited significant periods of 2.91\,h and 3.66\,h, which were not known in the literature. Furthermore, a period of 7.28\,h was found in both the He\,{\sc i + ii} $\lambda$4026 and He\,{\sc i} $\lambda$4922 lines, while the He\,{\sc i + ii} $\lambda$4026 and He\,{\sc ii} $\lambda$4200 lines showed periods of 5.37\,h and 5.15\,h, respectively. These periods are fairly close to the previously detected 5.34-h period, which was suggested to be a low-angular degree non-radial pulsation period by \citet{kambe1997multiperiodicity}. Therefore, our analysis demonstrates not only that pulsation periods can be detected in the $v\,\sin\,i$ time series using the Fourier method, but that the scatter in $v\,\sin\,i$ for different spectral lines depends on the pulsation properties of the star. Finally, our results highlight the sensitivity of the Fourier method to measuring $v\,\sin\,i$ in the presence of LPV caused by pulsations in massive stars.

\subsection{Macroturbulence}

\begin{figure*}
    \centering
    \includegraphics[width=0.99\textwidth]{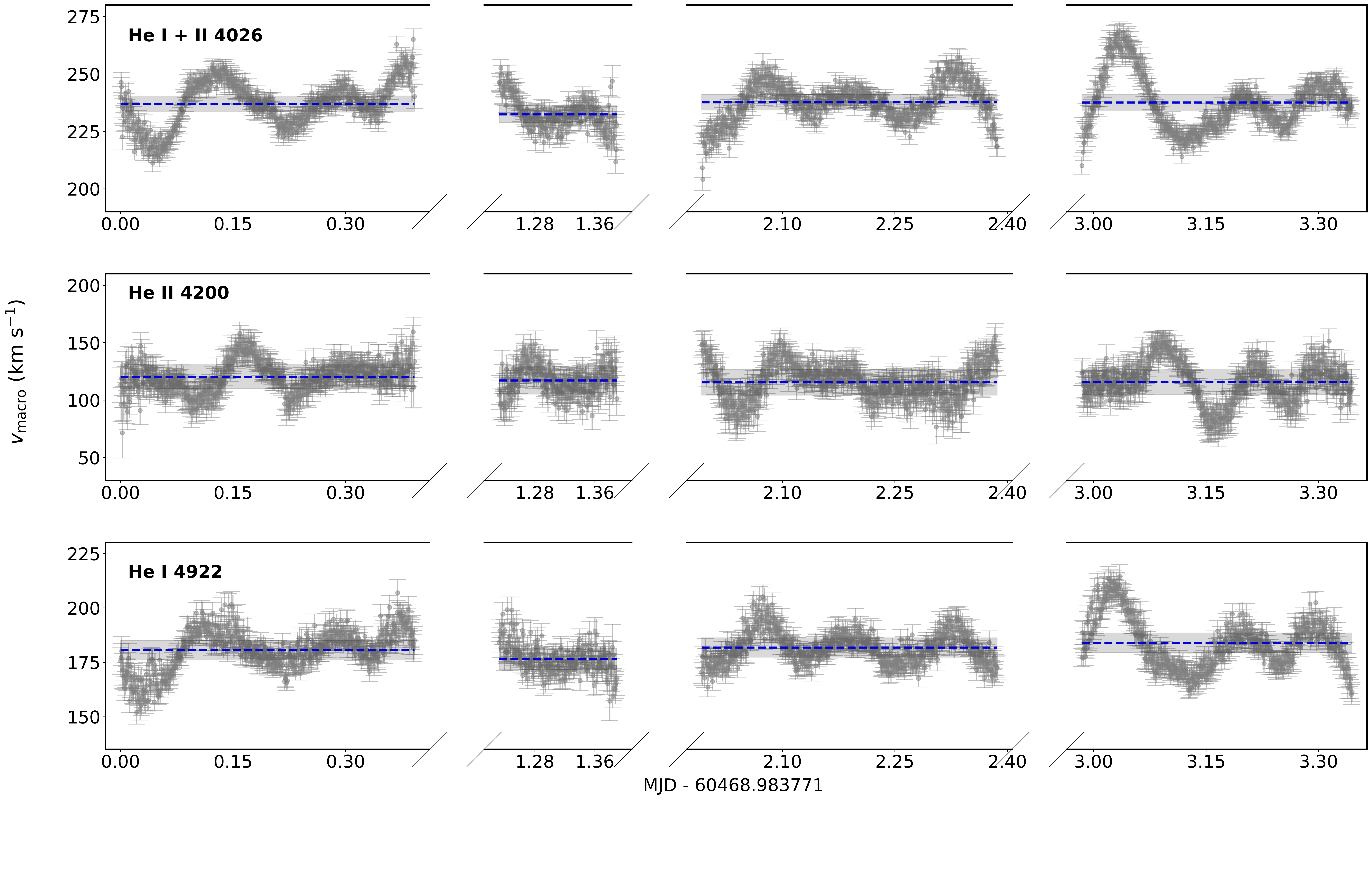}
    \includegraphics[width=0.99\textwidth]{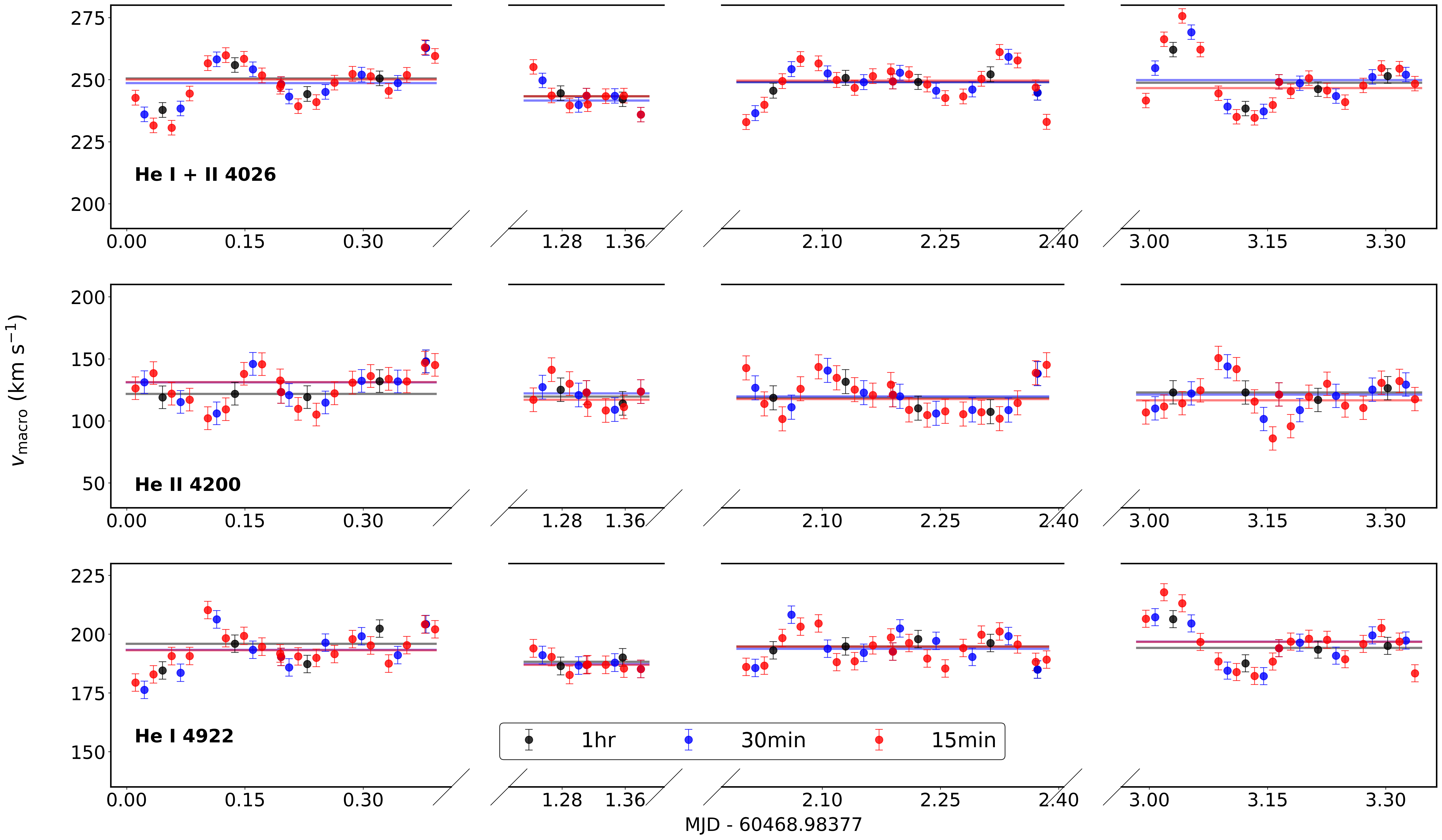}
    \caption{Macroturbulent velocity (i.e. $v_{\rm macro}$) for three helium lines in $\zeta$~Oph, with an x-axis normalised to the time stamp of the first spectrum (i.e. MJD = 60468.983771). Top panel: time series of all 1706 spectra with 30-s exposure times (in grey), which shows a peak-to-peak variability in $v_{\rm macro}$ of up to 88~km\,s$^{-1}$. Blue dashed lines are the mean of the $v_{\rm macro}$ values for each night, which are consistent within their uncertainties as indicated by the shaded region. Bottom panel: $v_{\rm macro}$ time series for stacked spectra emulating 1-h (black), 30-min (blue) and 15-min (red) integration times. The peak-to-peak variability is diminished for longer integration times, but the mean $v_{\rm macro}$ is larger for longer integration times, with the nightly mean values denoted as solid lines.}
    \label{vmacro_timeseries}
\end{figure*}

We estimate the macroturbulent velocity broadening (i.e. $v_{\rm macro}$) for the three helium lines by assuming a constant rotation rate fixed at 327~km\,s$^{-1}$, which is the mean of the $v\,\sin\,i$ values for all three helium lines, as derived in Section~\ref{sec:rotation} and shown in Fig.~\ref{vsini_time-series}. We take the standard approach to constrain macroturbulence in $\zeta$~Oph by following the method of \citet{simon2007fourier}, which assumes that the intrinsic line profile is Gaussian, a full-width-at-half-maximum determined by the spectral resolving power, and a line depth set by the observed equivalent width of the line profile. The overall spectral line profile is fitted by including two broadening components: (i) rotation and (ii) $v_{\rm macro}$, using the isotropic radial-tangential model from \citet{gray2005observation}. In this isotropic model of macroturbulence, equal fractions of the stellar surface contribute to radial and tangential motion. Whilst this may not be physical, it is, however, a common assumption for massive stars \citep{ryans_macroturbulence_in_BSGs_2002, simon2010observational, simon_iacob_2017}. Following the formalism of \citet{gray2005observation}, the macroturbulent broadening profile is given by
\begin{equation}
M(\Delta\lambda) = \frac{2 A_{R}\Delta\lambda}{\sqrt{\pi}\zeta^{2}_{R}}\int^{\zeta_{R}/\Delta\lambda}_{0}e^{-1/u^{2}}\,{\rm d}u + \frac{2 A_{T}\Delta\lambda}{\sqrt{\pi}\zeta^{2}_{T}}\int^{\zeta_{T}/\Delta\lambda}_{0}e^{-1/u^{2}}\,{\rm d}u,
\label{eq:macro_profile}
\end{equation}
\noindent where $\zeta_{R}$ and $\zeta_{T}$ represent the radial and tangential velocity amplitudes, and $A_R$ and $A_T$ are the fractional flux contributions from the radial and tangential components, respectively.

We have intentionally adopted an isotropic radial-tangential model for $v_{\rm macro}$ for our analysis of $\zeta$~Oph, which assumes equal flux contributions (i.e. $A_R = A_T = 0.5$) and equal amplitudes for the radial and tangential velocity components (i.e. $\zeta_{R} = \zeta_{T} = v_{\rm macro}$), since this is commonly implemented in spectroscopic studies of massive stars. While these assumptions may not strictly hold, disentangling both the flux ratio and the velocity components from spectroscopy alone is inherently degenerate and necessitates simplifications (e.g. \citealt{nadya_uver_feros_spectroscopy}). Moreover, the primary aim here is to demonstrate how common approaches of inferring macroturbulent velocities are significantly affected by LPV caused by pulsations on short time scales. For simplicity, we have also fixed the $v\,\sin\,i$ as mentioned previously, and do not consider the effect of limb darkening or gravity darkening.

We applied this isotropic radial-tangential model to each individual ESPRESSO spectrum of $\zeta$~Oph with a 30-s exposure time. Figure~\ref{vmacro_timeseries} shows the $v_{\rm macro}$ time series for the three helium lines considered in this work. The $v_{\rm macro}$ measurements from individual 30-s exposures show significant peak-to-peak scatter, up to 88~km\,s$^{-1}$, which is much larger than the variability seen in RV or $v\,\sin\,i$ (c.f Figs.~\ref{rv_timeseries} and \ref{vsini_time-series}). This clearly indicates that macroturbulence is not a static velocity field, but instead exhibits large-amplitude and short-period variability, which has a specific average value and different temporal behaviour for each spectral line.

We also applied the isotropic radial-tangential model to the time-series of stacked spectra that emulate longer exposure times. We do this to investigate potential biases in inferring macroturbulence when integration times are longer and thus more likely to suffer from pulsation-induced smearing of the spectral line profile. Such longer integration times of order tens of minutes are typically needed in high-resolution spectroscopic studies of fainter massive stars. We show the $v_{\rm macro}$ values derived from stacked spectra emulating 15-min, 30-min, and 1-h integration times in the lower panels of Fig.~\ref{vmacro_timeseries}. Importantly, we find that the mean $v_{\rm macro}$ values for longer integration times are systematically larger, which is due to the pulsational smearing of the line profile for longer exposures, changing the total equivalent width of the line profile.

\begin{figure}
    \centering
    \includegraphics[width=0.99\columnwidth]{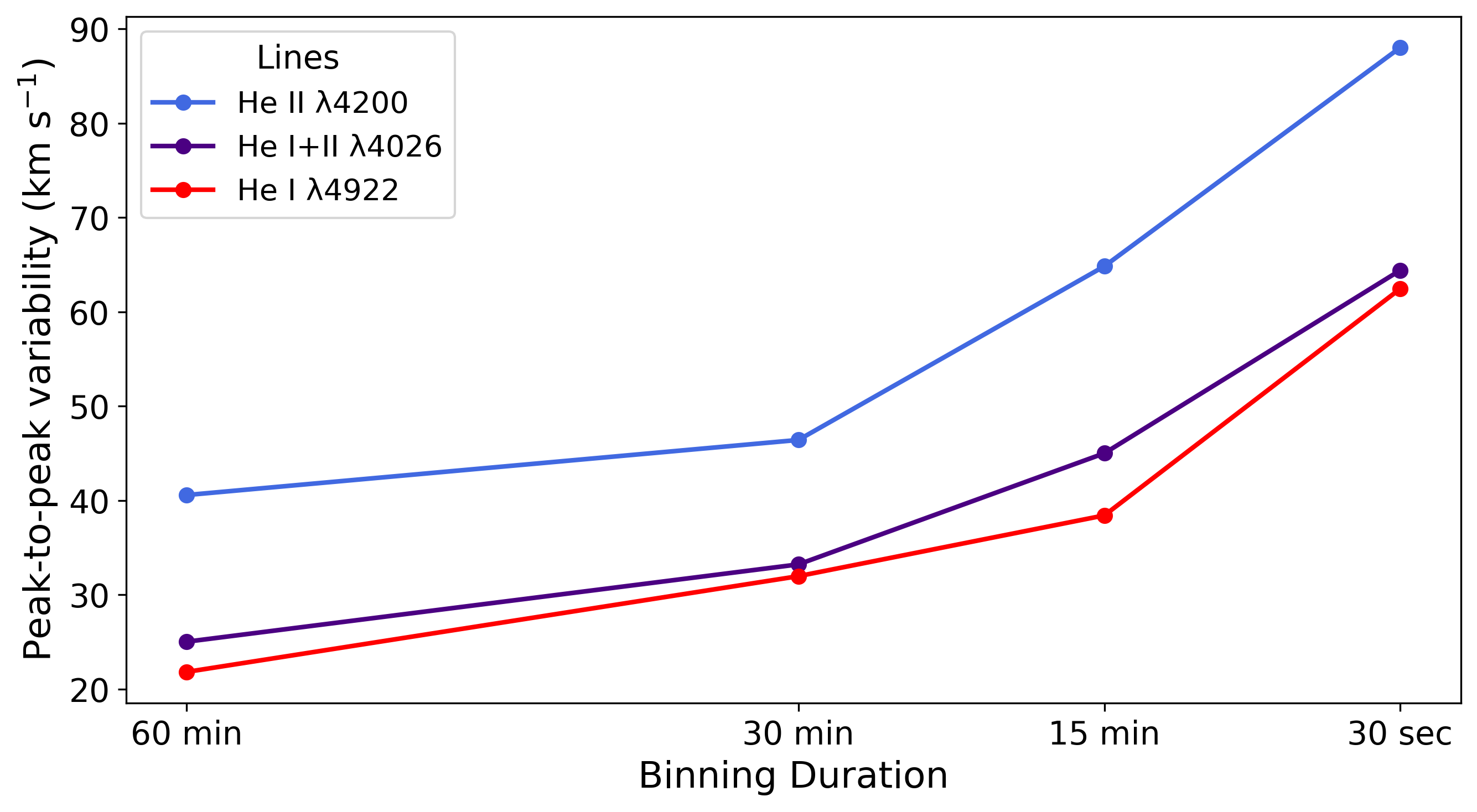}
    \caption{Scatter in the peak-to-peak variability of $v_{\rm macro}$ for the three helium lines for the different sets of binned spectra.}
    \label{variability_in_bins}
\end{figure}

Moreover, our analysis demonstrates that the peak-to-peak scatter in the $v_{\rm macro}$ values is line dependent. For the He\,{\sc ii} $\lambda$4200 line, the peak-to-peak scatter is systematically larger for shorter integration times, increasing from 40~km\,s$^{-1}$ for the 1-h stacked spectra to 88~km\,s$^{-1}$ for the individual spectra with 30-s exposure times. For the other two helium lines, the scatter is approximately 20~km\,s$^{-1}$ for the 1-h stacked spectra, and exceeds 60~km\,s$^{-1}$ for the individual spectra with 30-s exposure times. In Fig.~\ref{variability_in_bins} we summarise the scatter in $v_{\rm macro}$ as a function of different time-binning, and show that it is systematically larger for the shorter integration times. This implies that high-resolution spectroscopy of fainter massive stars, for which longer integration times are typically required, leads to biased estimates of both the amplitude and variability of macroturbulence. Specifically, the mean macroturbulence may be overestimated due to pulsational smearing, while its intrinsic variability may be underestimated.

We applied our iterative pre-whitening method to the $v_{\rm macro}$ time series for the individual spectra with 30-s exposure times and found that all three helium lines exhibit multiperiodic variability. The significant periods of 1.39\,h, 2.91\,h, 3.33\,h, 4.63\,h, and 5.34\,h were observed in all three helium lines. As previously mentioned, the 3.33-h period is a known non-radial pulsation period with spherical harmonic geometry of $\ell = |m| = 4$ \citep{reid1993time, kambe1997multiperiodicity}. The 5.34-h period was also observed by \citet{kambe1997multiperiodicity}. Based on the fact that it was only dominant in the line wings and its phase decreased slightly towards longer wavelength,  \citet{kambe1997multiperiodicity} suggested that this period is likely due to low-angular degree non-radial pulsation. The 4.63-h period is also likely to be either a radial first overtone or a low-spherical order non-radial pulsation, as reported by \citet{walker2005pulsations}.

Additionally, the He\,{\sc i + ii} $\lambda$4026 and He\,{\sc ii} $\lambda$4200 lines show periods of 2.43\,h and 2.44\,h, respectively. These periods are consistent with the previously reported 2.43-h period by \citet{kambe_short_term_lpv1993} and \citet{reid1993time}. But, whether this period is a true period of a non-radial pulsation mode or an alias of the reported 2.01-h period is debated in the literature \citep{ kambe1997multiperiodicity, balona1999moving}. However, in our analysis, we observed significant periods of 2.04\,h and 2.15\,h in the He\,{\sc i + ii} $\lambda$4026 and He\,{\sc ii} $\lambda$4200 lines, respectively. Within their uncertainty, both these periods are consistent with the 2.01-h period, which is suggested to be a non-radial pulsation with spherical geometry $\ell =7-8$ by \citet{kambe1997multiperiodicity} and \citet{balona1999moving}. In summary, our results strongly support that the large-scale and short-period variability of macroturbulence in $\zeta$~Oph is caused by multiperiodic pulsations.

\section{Impact of variable macroturbulence on inferred atmospheric parameters}

\begin{figure}
    \centering
    \includegraphics[width=0.99\columnwidth]{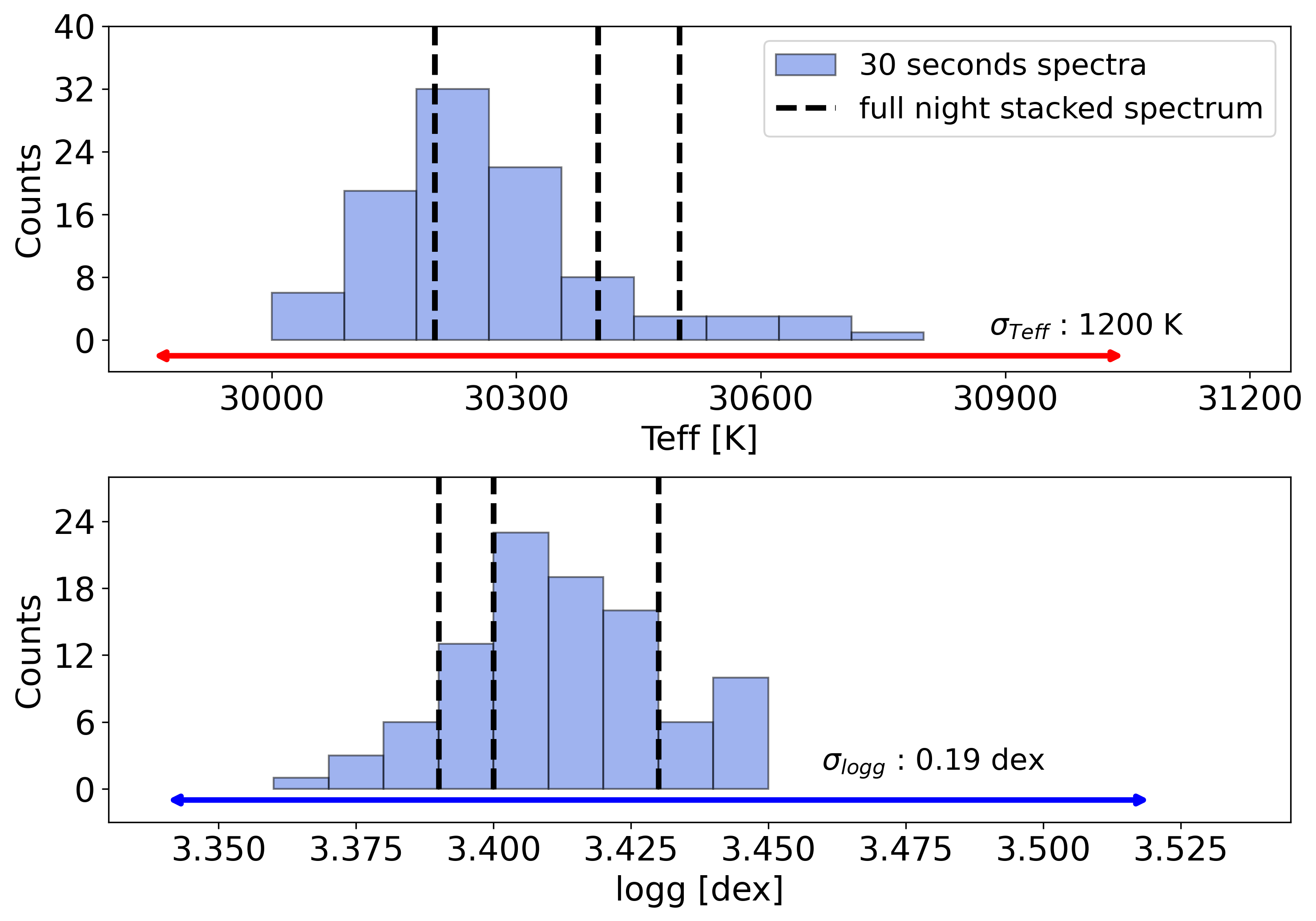}
    \caption{Histograms of $T_{\rm eff}$ and $\log\,g$ for the 30-s spectra. The vertical dashed lines in black indicate the best-fit values obtained from the full-night stacked spectra. The red and blue arrows indicate the average confidence interval on the $T_{\rm eff}$ and $\log\,g$ estimates from individual spectra.}
    \label{stellar_params}
\end{figure}

Having established the significant large-scale and short-period multiperiodic variability in macroturbulence for $\zeta$~Oph, which is much larger than that of the RV and $v\,\sin\,i$ time series, we also investigated the impact this has on inference of stellar parameters. We use the grid-based quantitative spectroscopy tool {\sc IACOB-gbat} \citep{iacob_gbat}, which employs {\sc FASTWIND} \citep{santolaya_fastwind_paper_1997, fastwind_jo_puls_2005, jon_and_puls_FASTWIND_2018, fastwind_updates_2020} atmospheric models to derive $T_{\rm eff}$ and $\log\,g$ for subsets of spectra that sample the full range in $v_{\rm macro}$ as shown in Fig.~\ref{vmacro_timeseries}. In this way, the full range of $v_{\rm macro}$ values are propagated to create distributions for $T_{\rm eff}$ and $\log\,g$ and test by how much these parameters vary because of variable macroturbulence. 

Figure~\ref{stellar_params} shows histograms of the derived $T_{\rm eff}$ and $\log\,g$ estimates for a subset of 100 spectra with 30-s exposure times. We find a non-negligible scatter in the atmospheric parameters caused by variable macroturbulence, with a range almost as large as the formal uncertainties for each parameter when considering a single spectrum. This demonstrates that while time-dependent macroturbulence contributes to systematic and random scatter in the inferred stellar parameters, such effects are only detectable with the ultra-high precision offered by high-resolution spectrographs such as ESPRESSO for the brightest massive stars.

\section{Comparison to archival and TESS photometry}\label{sec:TESS_photometry}

\begin{figure*}
    \centering
    \includegraphics[width=0.99\textwidth]{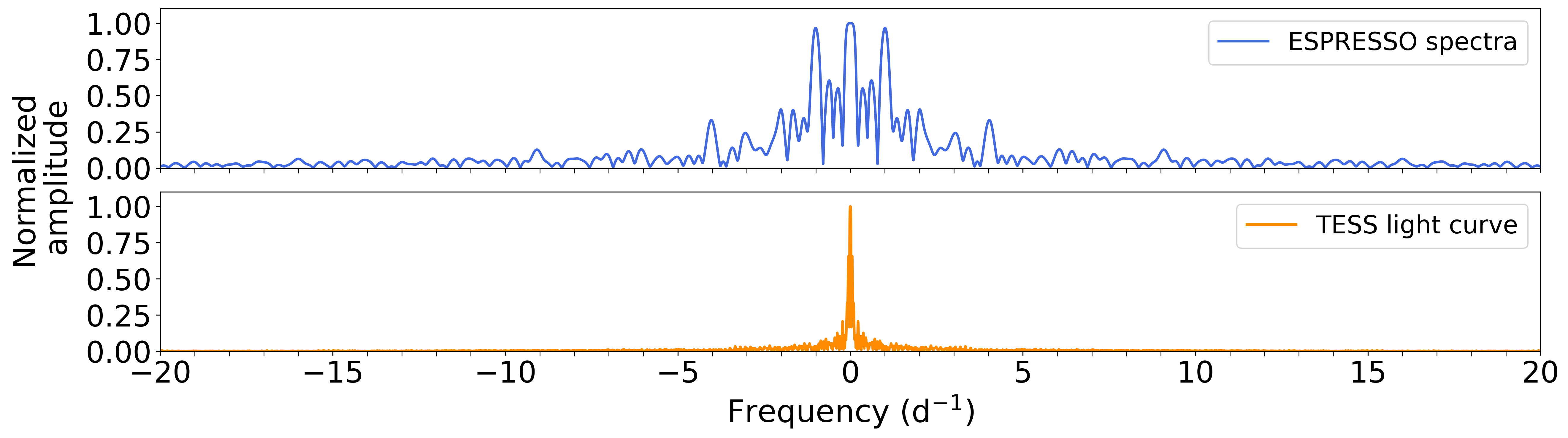}
    \caption{Spectral windows for the ESPRESSO spectroscopic time series (top panel) and the TESS photometric data (bottom panel) analysed in this work.}
    \label{spec_window}
\end{figure*}

\begin{figure*}
    \centering
    \includegraphics[width=0.99\textwidth]{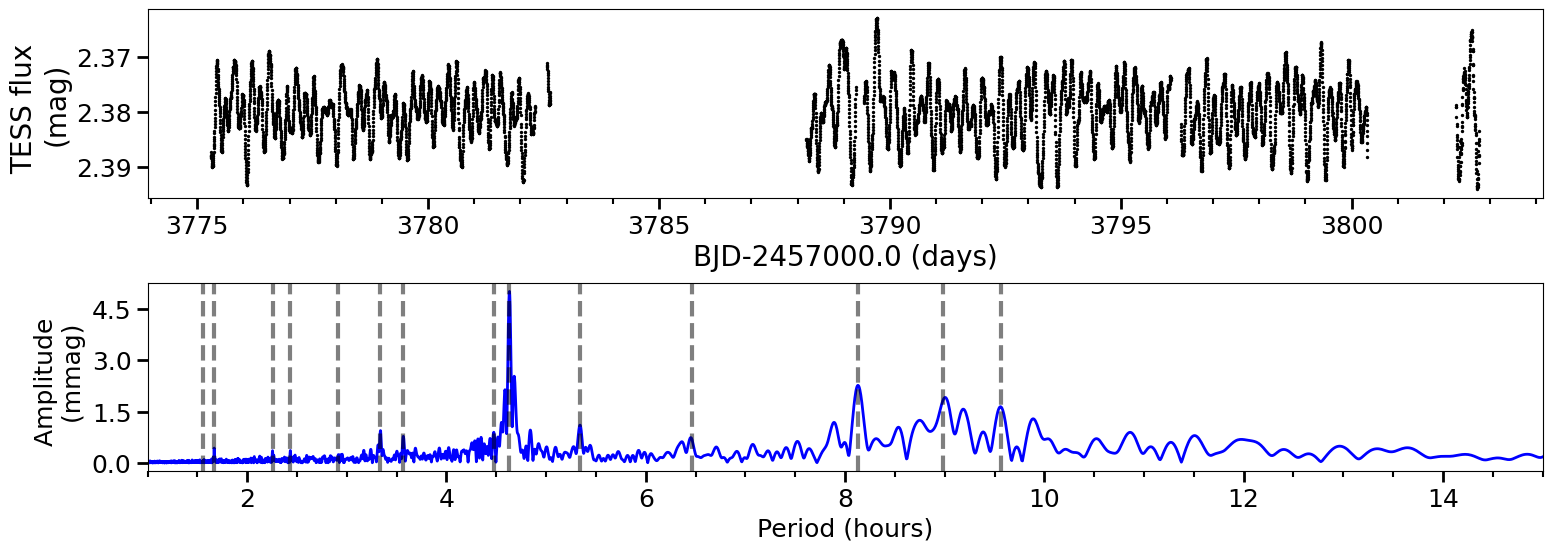}
    \caption{Top panel: Extracted TESS light curve from sector~91 of $\zeta$~Oph extracted using simple aperture photometry. Bottom panel: Lomb-Scargle periodogram of the TESS light curve, with significant periods identified in this work shown as dashed grey lines.}
    \label{fig:TESS}
\end{figure*}

\begin{figure*}
    \centering
    \includegraphics[width=0.99\textwidth]{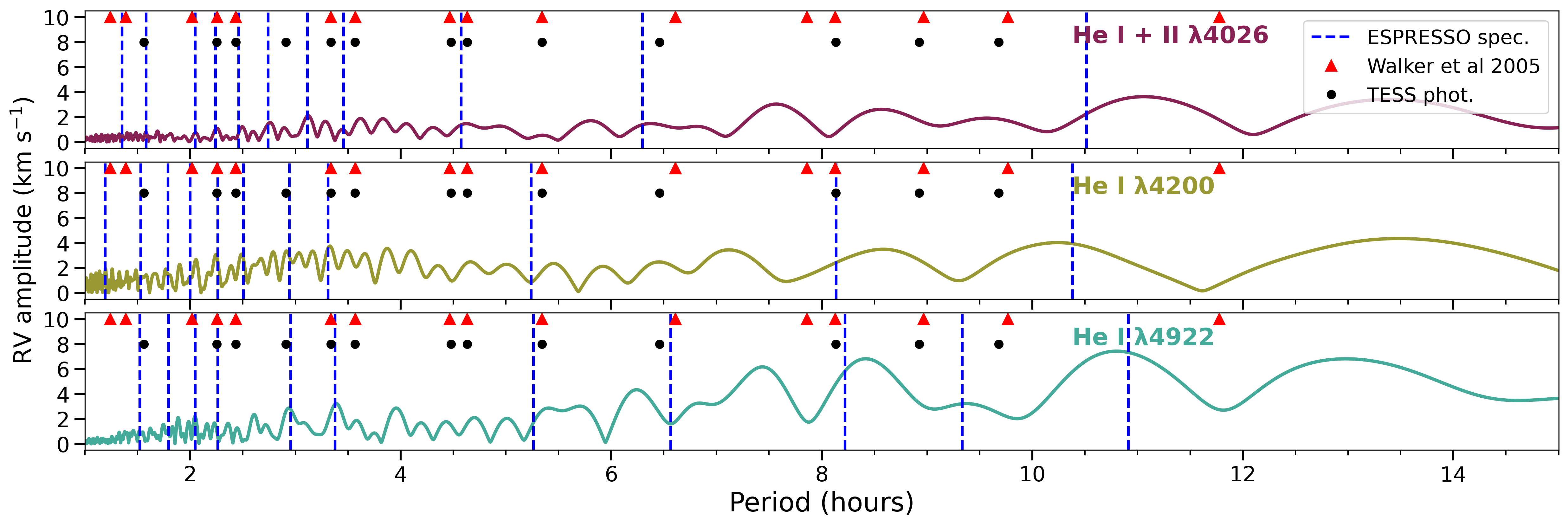}
    \includegraphics[width=0.99\textwidth]{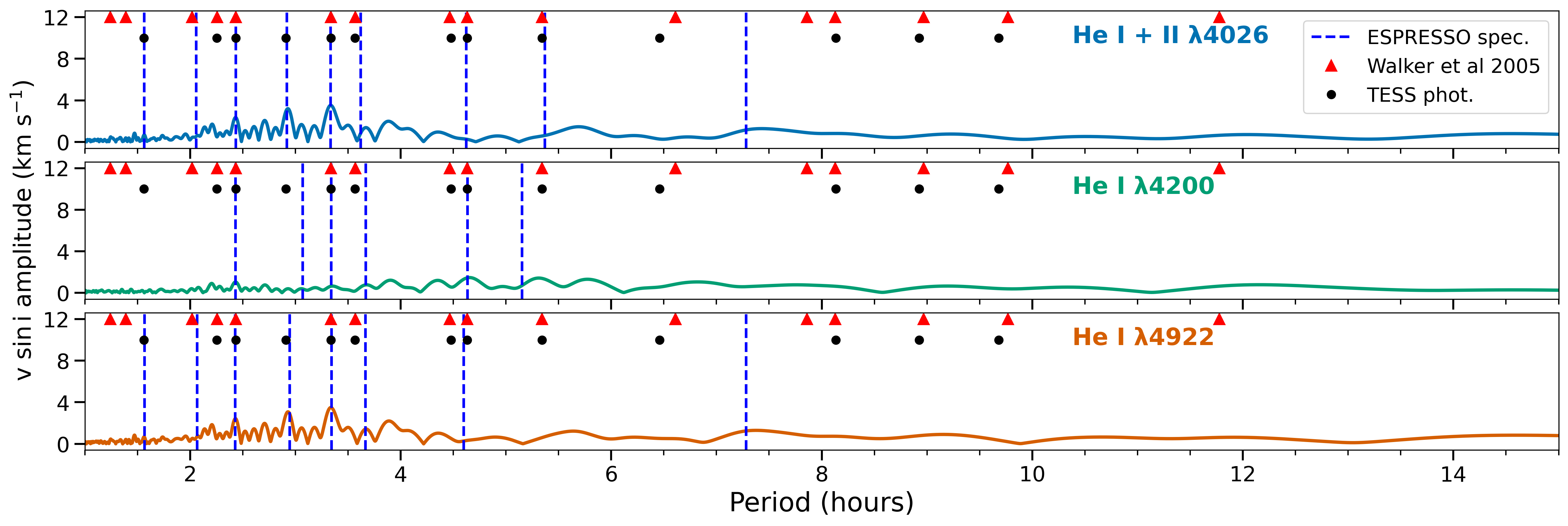}
    \includegraphics[width=0.99\textwidth]{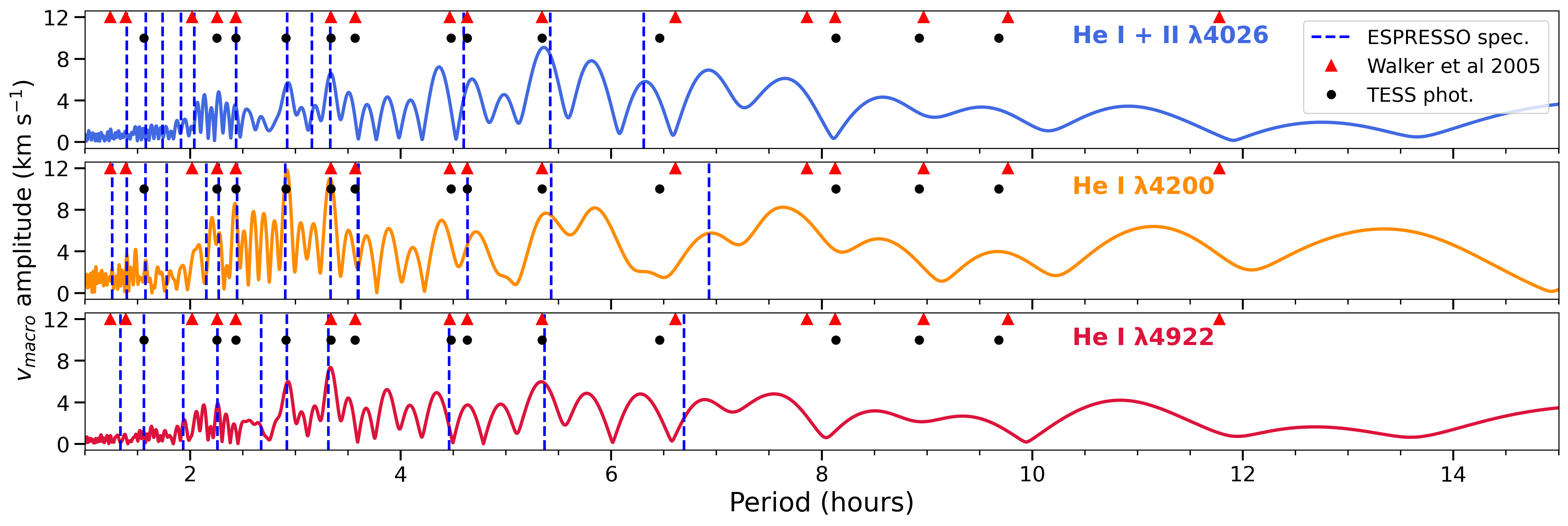}
    \caption{From top to bottom, Lomb-Scargle periodograms for the RV, $v\,\sin\,i$ and $v_{\rm macro}$ time series for the same three helium lines. In each panel, the blue-dashed lines indicate the significant periods extracted for each spectroscopic time-series using the pre-whitening method described in Section~\ref{section:lpv_analysis}, the black circles indicate the pulsation periods extracted from the TESS light curve, and the red triangles correspond to the pulsation periods from ground-based and MOST photometry reported by \citet{walker2005pulsations}.}
    \label{ft_spectra}
\end{figure*}

$\zeta$~Oph has been studied extensively using ground-based photometry as well as space-based photometry from the MOST, SMEI, and WIRE missions. Several independent analyses have revealed multiperiodic pulsations with quasi-periodic amplitudes and frequencies \citep[e.g.][]{balona1999moving, walker2005pulsations, howarth2014amplitude}. All photometric studies consistently report a dominant pulsation period of 4.63\,h. Using MOST photometry, \citet{walker2005pulsations} also detected the spectroscopically identified periods of 2.43\,h, 3.33\,h, and 5.34\,h, but did not recover the 2.01\,h period reported previously by \citet{kambe1997multiperiodicity}. Following the approach of \citet{telting_NRP_amplitude_phase_models_1997} and using simultaneous photometry and spectroscopy, \citet{walker2005pulsations} suggested that the 4.63-h pulsation period is a non-radial mode with $\ell \in \{3,4\}$, while the 2.43-h and 3.33-h periods correspond to $\ell = 7$ and $\ell \in \{3,4\}$, respectively. However, using non-adiabatic pulsation analysis for evolutionary models of a 23-M$_\odot$ star, \citet{walker2005pulsations} found that the 4.63-h period may also be consistent with a radial first overtone or a non-radial mode with $(\ell, m)$ = (1, -1) or (3, 1). 

The disagreement in the inferred mode geometry between different studies, or between methods within the same study, may arise because of coupling between rotation and pulsation modes in $\zeta$~Oph, which means that the geometry of a pulsation mode cannot be described by a single spherical
harmonic. For example, coupling between different spherical harmonics, such as between spheroidal and toroidal components, can occur (see \citealt{telting_NRP_amplitude_phase_models_1997}). This can lead to different inferred $\ell$-values depending on whether such coupling is considered or not when modelling spectroscopic or photometric variability (see e.g. \citealt{Daszynska2002}). Therefore, whilst it is clear that $\zeta$~Oph is multi-periodic pulsator, there is still ambiguity in the spherical harmonic geometries of its pulsation periods.

To complement our spectroscopic analysis of $\zeta$~Oph, we also analyse recently assembled TESS space photometry. The NASA TESS mission is an all-sky survey providing optical light curves at high cadence and had a nominal mission to identify Earth-like planets transiting bright stars \citep{Ricker2015}. TESS observed $\zeta$~Oph for the first time in its sector~91 in April 2025. Despite a minor safe-mode event, TESS assembled 28-d time-series photometry with a cadence of 200~s in its full-frame images (FFIs). We used the {\sc lightkurve} software package \citep{lightkurve2018} to take a cutout of the TESS FFIs for sector~91 and followed the methodology of \citet{bowman2022cubespec} to select an optimum aperture mask to extract a light curve. We then used principal component analysis (PCA) to detrend the light curve of remaining instrumental trends. Since $\zeta$~Oph is a bright star, it is saturated in the TESS FFIs, but we can capture sufficient flux by including the bleed columns within our aperture mask, which is shown in Fig.~\ref{aperture}. The spectral window of the extracted TESS light curve is shown alongside that of the ESPRESSO time series in Fig.~\ref{spec_window}, which demonstrates the improved nature of aliasing for near-continuous space photometry compared to ground-based spectroscopy.

The detrended sector 91 TESS light curve and its corresponding Lomb-Scargle periodogram are shown in Fig.~\ref{fig:TESS}. We applied an iterative pre-whitening method as before to detect the significant periods from the TESS data set, which are indicated in Fig.~\ref{fig:TESS} and included in Table~\ref{tab:frequencies} along with their least-squares fitting uncertainties. These new TESS data allow us to recover several significant pulsation periods that have been previously reported in ground-based photometric studies by \citet{balona1999moving} and \citet{walker2005pulsations}, but also detect several new significant pulsation periods.

To compare the significant periods found in the TESS light curve and the spectroscopic RV, $v\,\sin\,i$, $v_{\rm macro}$ time series, we show the Lomb-Scargle periodograms of each spectroscopic data set in Fig.~\ref{ft_spectra}. For ease of comparison between the data sets, significant periods from the TESS light curve are marked by black circles, while periods from archival photometry \citep{walker2005pulsations} are marked with red triangles, and significant periods from each spectroscopic time series are shown as vertical dashed blue lines. Among these, the previously known pulsation periods of 3.33\,h and 2.43\,h are consistently detected across all spectroscopic data sets and are also clearly present in the TESS light curve (also see \citealt{kambe_short_term_lpv1993, kambe1997multiperiodicity, balona1999moving}). A period of 2.91\,h is similarly detected in both photometric and spectroscopic data, suggesting that it may represent a genuine pulsation frequency, although this period has never been reported before in the literature.

Additionally, the 4.63-h period, which is significant in both the $v\,\sin\,i$ and $v_{\rm macro}$ time series, is also identified in the TESS photometry with the highest S/N of all pulsation periods. This periodicity was initially reported by \citet{balona1999moving} based on ground-based observations from 1985 but was notably absent in subsequent photometric campaigns. However, \citet{walker2005pulsations} and \citet{howarth2014amplitude} again reported the detection of the 4.63-h period based on the MOST, SMEI and WIRE photometry from 2003 to 2008. Its detection in the TESS light curve with high significance strongly supports the interpretation that the 4.63-h period corresponds to a pulsation period rather than an alias.

The 5.34-h pulsation period reported by \citet{kambe1997multiperiodicity} and  \citet{walker2005pulsations} is also recovered in the TESS data. We also detect significant periods of 5.42\,h, 5.43\,h and 5.36\,h in the $v_{\rm macro}$ time series of the three diagnostic helium lines from our ESPRESSO spectra, all of which are consistent with the 5.34-h period within their uncertainties. The 2.01-h pulsation period is not detected in the TESS photometry, despite being consistently observed in our RV time series for all three helium lines, as well as in the $v\,\sin\,i$ and $v_{\rm macro}$ time series for the He\,{\sc i +\sc ii} $\lambda$4026 line and the $v\,\sin\,i$ time series of the He\,{\sc i} $\lambda$4922 line. Notably, the sum of the frequencies corresponding to the periods of 4.63\,h ($f_1$) and 3.56\,h ($f_2$) (i.e. a combination frequency of $f_1 + f_2$) yields a frequency at 11.93~d$^{-1}$ in the TESS data, which is very close to the 11.98~d$^{-1}$ frequency associated with the reported alias period of 2.01\,h. This supports the interpretation that the 2.01-h period is a combination frequency, while the 2.43\,h period is the real pulsation period rather than an alias of the 2.01-h period.

While several other longer periods known from archival photometry were also observed in the TESS data, these were not observed in the ESPRESSO data. However, this is because the ESPRESSO spectroscopic time series has a significantly different temporal sampling and spectral window, resulting in reduced sensitivity at low frequencies. For example, the time span of our ESPRESSO data is four consecutive nights, meaning that periods longer than several days are subject to aliasing. Consequently, these low-frequency signals cannot be identified with confidence in the spectroscopic diagnostics.

\section{Discussion and conclusions}\label{sec:discussion}

Using high-cadence, high-resolution, and high-S/N ESPRESSO spectroscopy, we provide compelling evidence that macroturbulence in the rapidly rotating and pulsating massive star $\zeta$~Oph is not a static velocity field but instead exhibits large-amplitude, short-period, and line-specific variability caused by pulsations. The $v_{\rm macro}$ time series in particular derived from individual 30-s spectra shows peak-to-peak scatter of up to 88~km\,s$^{-1}$ and periods of order hours. Furthermore, using different stacked time series to emulate longer integration times, we demonstrate that the inferred amplitude of the variability in macroturbulence depends on the integration time of the spectra because of pulsation-induced smearing of the line profile. For spectra with integration times comparable to the shortest pulsation period (i.e. $\sim$1\,h), the observed scatter in $v_{\rm macro}$ is significantly reduced down to $\sim$20~km\,s$^{-1}$. Moreover, the mean $v_{\rm macro}$ values inferred from longer integration times are systematically higher than those derived from shorter integration time spectra. This is due to the pulsational smearing of the line profile in long-exposure observations. This result is similar to previous works by \citet{kambe1997multiperiodicity} and \citet{aerts2009collective} who looked at the artificial broadening of spectral lines caused by pulsations in massive stars. This also suggests that traditional approaches relying on a single stacked spectrum systematically overestimate macroturbulent broadening.

However, there are limitations to our analysis. The helium lines used in the analysis do have small amounts of Stark broadening, hence our study of macroturbulent velocities are slightly overestimated. However, Stark broadening is much smaller than the total non-rotational additional broadening we refer to as macroturbulence. Additionally, the assumption of isotropic macroturbulence does not fully hold when the observational cadence is much shorter than the dominant pulsation period. In such cases, pulsations can produce asymmetric line profiles, and the symmetric isotropic macroturbulence may not be an accurate representation of the line profile. This effect is relatively small compared to the dominant rotational broadening for rapidly rotating stars such as $\zeta$~Oph, but remains non-negligible. Regardless, our study has concretely demonstrated how conventional methods of inferring macroturbulence can be biased by pulsations and instrumental setup. As future work, we will model the pulsational effects in more detail by incorporating line asymmetry and by accounting for the differing contributions of radial and tangential velocity components from pulsations. Disentangling the ratio of radial to tangential velocities is particularly important, as gravity modes typically have $\zeta_t \gg \zeta_r$, which may differ from other instabilities such as convection.

We assessed the impact of time-dependent macroturbulence on stellar parameter inference by applying the grid-based fitting tool {\sc IACOB-gbat} to spectra sampling the full range of observed $v_{\rm macro}$ values. This approach enabled us to construct distributions for $T_{\rm eff}$ and $\log g$. The scatter in the derived parameters is comparable to the formal uncertainties on an individual spectrum and indicates that macroturbulence variability can introduce both systematic and random errors in parameter estimation. However, these effects are only apparent with the ultra-high signal-to-noise and time resolution of ESPRESSO and would likely remain undetected in conventional, time-averaged datasets.

Iterative pre-whitening applied to the RV, $v\,\sin\,i$, and $v_{\rm macro}$ time series revealed a large number of significant periods, which reinforces the fact that $\zeta$~Oph is a multi-periodic pulsator with significant LPV. Using the first TESS light curve ever assembled for $\zeta$~Oph, we also identified the significant pulsation periods, many of which are in common with our ESPRESSO spectroscopic time series. The 2.43\,h, 2.91\,h, 3.33\,h and 4.63\, periods were observed consistently in both the $v\,\sin\,i$ and $v_{\rm macro}$ time series across all three diagnostic helium lines, and in our TESS light curve. Previous spectroscopic investigations by \citet{kambe1990spectroscopic, kambe_short_term_lpv1993, kambe1997multiperiodicity} and \citet{balona1999moving}, along with photometric follow-up studies by \citet{walker2005pulsations} and \citet{howarth2014amplitude}, provided strong evidence that the 3.33\,h and 2.43\,h periods are caused by non-radial pulsations. A period of 5.34\,h was found in both the TESS light curve and the $v_{\rm macro}$ time series, which was also found in previous studies and suggested to be a low-angular degree non-radial pulsation \citep{kambe1997multiperiodicity}. However, there remains some debate in the literature as to which of the 2.01\,h or 2.43\,h periods is an alias. In our analysis, the 2.43\,h period is clearly detected in the TESS photometry, while the 2.01\,h signal is not, thus we conclude that the 2.01\,h period is the alias owing to TESS's excellent duty cycle and spectral window.

The similar significant periods in our spectroscopic data to those identified as pulsation modes in TESS and archival photometry provides compelling evidence that the observed variability in macroturbulence in $\zeta$~Oph is driven by non-radial pulsations. Thus, our proof-of-concept using ESPRESSO data of $\zeta$~Oph demonstrates that pulsations have a critical role to play in the physics of (time-dependent) macroturbulence in massive stars. The variability observed in $v_{\rm macro}$ is not only large in amplitude but also quasi-periodic and spectral-line specific, and we demonstrate how it may be overestimated for fainter massive stars that require long exposure times. However, there remains ambiguity in the identification of spherical harmonic geometry for $\zeta$~Oph's pulsation modes, which is the subject of our future work. In the future, incorporating significant frequencies identified from TESS and archival photometry into physically motivated, time-dependent, and anisotropic prescriptions for macroturbulence will be instrumental in further improving our understanding of the structures of massive stars.

\begin{acknowledgements}

The authors gratefully acknowledge UK Research and Innovation (UKRI) in the form of a Frontier Research grant under the UK government's ERC Horizon Europe funding guarantee (SYMPHONY; PI Bowman; grant number: EP/Y031059/1), and a Royal Society University Research Fellowship (PI Bowman; grant number: URF{\textbackslash}R1{\textbackslash}231631).
This project received the support from the ``La Caixa'' Foundation (ID 100010434) under the fellowship code LCF/BQ/PI23/11970035.
SS-D acknowledges support from the State Research Agency (AEI) of the Spanish Ministry of Science and Innovation (MICIN) and the European Regional Development Fund, FEDER under grants LOS M{\'U}LTIPLES CANALES DE EVOLUCI{\'O}N TEMPRANA DE LAS ESTRELLAS MASIVAS PID2021-122397NB-C2.
Based on observations collected with the ESPRESSO spectrograph at the European Southern Observatory in Paranal, Chile, under ESO programme 113.26B9 (PI Bowman), which are accessible through the ESO archive (\url{https://www.eso.org/sso/login}).
Data products that support the results in this paper are publicly available via the Zenodo repository: \url{https://zenodo.org/records/TBD}.
\end{acknowledgements}

\bibliographystyle{aa}
\bibliography{example}
\begin{appendix}

\section{TESS photometry}\label{appendix:TESS}

\begin{figure}[h]
    \centering
    \includegraphics[width=0.99\textwidth]{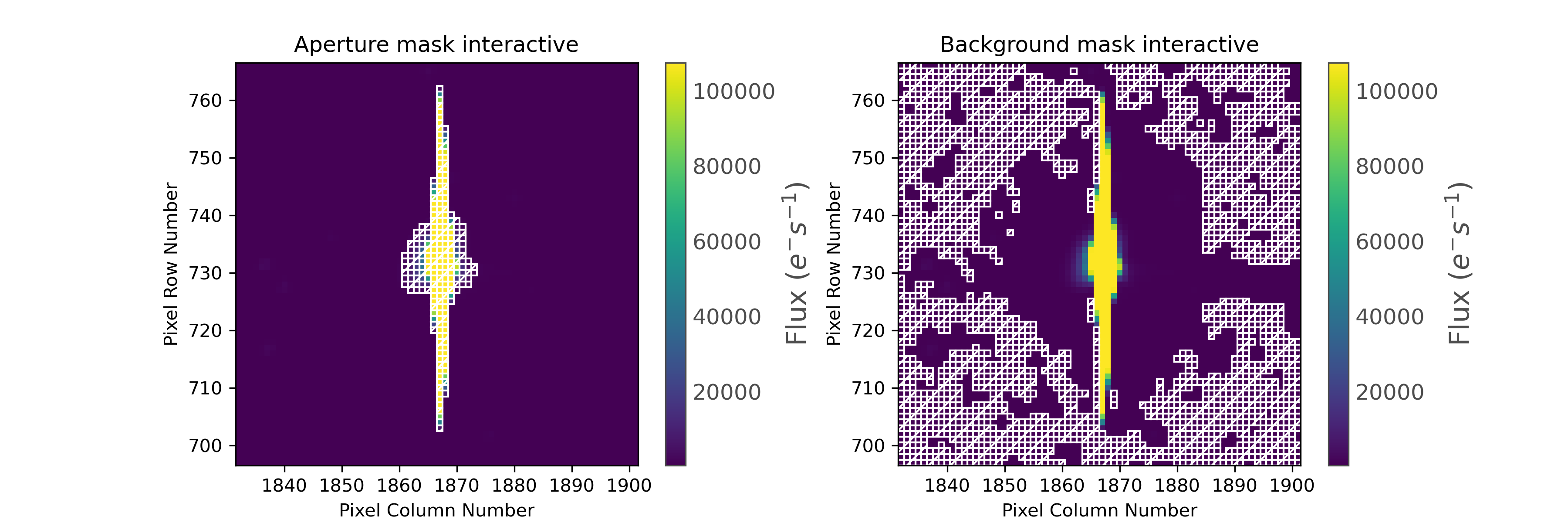}
    \caption{Our custom aperture and background masks based on the TESS FFI image of $\zeta$~Oph in sector 91.}
    \label{aperture}
\end{figure}

In Fig.~\ref{aperture}, we provide the TESS aperture and background masks used to extract the light curve in Fig.~\ref{fig:TESS}.

\FloatBarrier

\section{Extracted periodicity} \label{appendix:periods}

Significant periods identified across different time series, including from TESS and archival photometry, and ESPRESSO spectroscopy, and their respective uncertainties are provided in Table~\ref{tab:frequencies}.

\begin{sidewaystable*}
\small
\caption{Significant pulsation periods identified using archival photometry by \citet{walker2005pulsations} are shown in the first column, and those found in our frequency analysis of the $v_{\rm macro}$, $v\,\sin\,i$, RV, and TESS time series data sets in this work. All periods are given in the unit of hours.}
\label{tab:frequencies}
\centering
\begin{tabular}{c | c c c | c c c | c c c| c}
\hline\hline
Archival & \multicolumn{3}{c}{$v_{\rm macro}$ time series} & \multicolumn{3}{c}{$v\,\sin\,i$ time series} & \multicolumn{3}{c}{RV time series} & TESS \\
Photometry& He\,{\sc i+ii} & He\,{\sc ii} & He\,{\sc i} & He\,{\sc i+ii} & He\,{\sc ii} & He\,{\sc i} & He\,{\sc i+ii} & He\,{\sc ii} & He\,{\sc i} & Photometry \\
 & $\lambda$4026 & $\lambda$4200 & $\lambda$4922 & $\lambda$4026 & $\lambda$4200 & $\lambda$4922 &$\lambda$4026 & $\lambda$4200 & $\lambda$4922 \\
\hline
1.390  & 1.398 $\pm$ 0.057  & 1.399 $\pm$ 0.069 & 1.338 $\pm$ 0.050 & 1.564 $\pm$ 0.051 &                   & 1.567 $\pm$ 0.059  & 1.353 $\pm$ 0.107  & 1.196 $\pm$ 0.041 & 1.522 $\pm$0.055  &        \\
2.018  & 2.039 $\pm$ 0.365  & 2.153 $\pm$ 0.145 &                   & 2.060 $\pm$ 0.164 &                   & 2.068 $\pm$ 0.073  & 2.049 $\pm$ 0.268  & 2.002 $\pm$ 0.219 & 2.048 $\pm$ 0.077 &        \\
2.256  &                    & 2.273 $\pm$ 0.077 & 2.261 $\pm$ 0.064 &                   &                   &                    & 2.241 $\pm$ 0.168  & 2.263 $\pm$ 0.132 & 2.262 $\pm$ 0.624 & 2.2549 $\pm$ 0.2625\\
2.434  & 2.436 $\pm$ 0.104  & 2.446 $\pm$ 0.067 &                   & 2.435 $\pm$ 0.097 & 2.432 $\pm$ 0.073 & 2.428 $\pm$ 0.099  & 2.462 $\pm$ 0.559  & 2.506 $\pm$ 0.253 &                   & 2.4330 $\pm$ 0.1293\\
       & 2.921 $\pm$ 0.263  & 2.904 $\pm$ 0.077 & 2.919 $\pm$ 0.126 & 2.919 $\pm$ 0.132 & 3.071 $\pm$ 0.146 & 2.947 $\pm$ 0.081  & 2.742 $\pm$ 0.180  & 2.943 $\pm$ 0.381 & 2.954 $\pm$ 1.680 & 2.9091 $\pm$ 0.3232\\
3.337  & 3.331 $\pm$ 0.101  & 3.333 $\pm$ 0.185 & 3.312 $\pm$ 0.062 & 3.335 $\pm$ 0.069 & 3.340 $\pm$ 0.077 & 3.343 $\pm$ 0.069  & 3.116 $\pm$ 0.295  & 3.310 $\pm$ 0.232 & 3.378 $\pm$ 0.169 & 3.3374 $\pm$ 0.0944\\
3.571  &                    & 3.595 $\pm$ 0.624 &                   & 3.620 $\pm$ 0.290 & 3.668 $\pm$ 0.186 & 3.666 $\pm$ 0.181  & 3.457 $\pm$ 0.533  &                   &                   & 3.5671 $\pm$ 0.4703\\
4.633  & 4.599 $\pm$ 0.203  & 4.634 $\pm$ 0.284 & 4.459 $\pm$ 0.176 & 4.622 $\pm$ 0.262 & 4.635 $\pm$ 0.174 & 4.599 $\pm$ 0.189  & 4.576 $\pm$ 0.568  &                   &                   & 4.6318 $\pm$ 0.1074\\
5.344  & 5.421 $\pm$ 0.177  & 5.431 $\pm$ 0.315 & 5.367 $\pm$ 0.231 & 5.370 $\pm$ 0.285 & 5.153 $\pm$ 0.209 &                    & 6.296 $\pm$ 1.001  &                   & 5.262 $\pm$ 0.937 & 5.3413 $\pm$ 0.5233\\
6.610  & 6.311 $\pm$ 0.546  & 6.931 $\pm$ 0.957 & 6.693 $\pm$ 0.840 &                   &                   &                    &                    &                   & 6.566 $\pm$ 0.899 &        \\
7.880  &                    &                   &                   & 7.282 $\pm$ 0.665 &                   & 7.281 $\pm$ 0.458  &                    &                   &                   & 7.8751 $\pm$ 0.1823\\
8.127  &                    &                   &                   &                   &                   &                    &                    & 8.136 $\pm$ 1.138 & 8.220 $\pm$ 0.757 & 8.1302 $\pm$ 0.1739\\
9.770  &                    &                   &                   &                   &                   &                    &                    &                   &                   & 9.8652 $\pm$ 0.5292\\
11.777  &                    &                   &                   &                   &                   &                    &                    &                   &                   &        \\
15.625  &                    &                   &                   &                   &                   &                    &                    &                   &                   &        \\
\hline\hline 

\end{tabular}
\end{sidewaystable*}
\end{appendix}
\end{document}